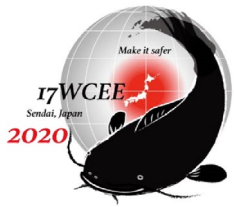

# UNCERTAINTY QUANTIFICATION OF STRUCTURAL SYSTEMS WITH SUBSET OF DATA

M.A. Hariri-Ardebili[(1)], F. Pourkamali-Anaraki[(2)], S. Sattar[(3)]

[(1)] *Research Associate, National Institute of Standards and Technology (NIST), amin.hariri@nist.gov*
[(2)] *Assistant Professor, University of Massachusetts Lowell, Farhad_Pourkamali@uml.edu*
[(3)] *Research Structural Engineer, National Institute of Standards and Technology (NIST), siamak.sattar@nist.gov*

## *Abstract*

Quantification of the impact of uncertainty in material properties as well as the input ground motion on structural responses is an important step in implementing a performance-based earthquake engineering (PBEE) framework. Among various sources of uncertainty, the variability in the input ground motions, a.k.a. record-to-record, greatly affects the assessment results. The objective of this paper is to quantify the uncertainty in structural response with hybrid uncertainty sources.

The outcome of the PBEE is commonly presented in terms of a fragility curve that shows probability of exceeding a specific damage state or demand with respect to an intensity measure of the input ground motion. Developing fragility curves usually requires conducting a large number of nonlinear time-history analyses. The fragility curve might be generated using wide-range probabilistic analysis methods such as incremental dynamic analysis, or cloud analysis. Fragility curves commonly capture the record-to-record variability, but the impact of material uncertainty is usually ignored as only one set of material properties is used each analysis. Propagating material uncertainties though the record-to-record variability requires a large number of analyses, which can become impractical for a large-scale structure. Therefore, there is a need to develop an alternative solution that reduces the required number of analysis while it does not adversely impact the accuracy of the result.

In this paper, multiple matrix completion methods are proposed and applied on a case study structure. The matrix completion method is a means to estimate the analyses results for the entire set of input parameters by conducting analysis for only a small subset of analyses. The main algorithmic contributions of our proposed method are twofold. First, we develop a sampling technique for choosing a subset of representative simulations, which allows improving the accuracy of the estimated response. An unsupervised machine learning technique is used for this purpose. Next, the proposed matrix completion method for uncertainty quantification is further refined by incorporating a regression model that is trained on the available partial simulations. The regression model improves the initial sampling as it provides a rough estimation of the structural responses.

Finally, the proposed algorithm is applied to a multi-degree-of-freedom system, and the structural responses (i.e., displacements and base shear) are estimated. Results show that the proposed algorithm can effectively estimate the response from a full set of nonlinear simulations by conducting analyses only on a small portion of the set.

## 1. Introduction

There are typically various sources of uncertainty in engineering problems. In a very broad classification, they can be categorized as aleatory or epistemic [1] uncertainties. The aleatory uncertainty is the natural randomness in a process. It is also called inherent randomness, stochastic, or variable uncertainty. This type of uncertainty cannot be reduced by performing more experiments or exhaustive measurements. The epistemic uncertainty is the scientific uncertainty in the model of the process, and is due to the limited data and knowledge. It is also called subjective uncertainty, and is reducible with improvements to the model and the related parameters.

The basic qualifier in aleatory and epistemic uncertainties refers to directly observable quantities such as material properties (e.g., strength and stiffness), loads (e.g., earthquake magnitude and sea wave height), environmental phenomena (e.g., temperature, alkali-aggregate reaction), and geometric dimensions (e.g., section size). Distinctions between aleatory and epistemic uncertainties are addressed in different engineering fields, e.g. structures and pipelines [2], groundwater flow [3], seismic analysis of reinforced concrete framed structures [4], buildings [5], concrete dams [6], flood frequency analysis [7], and coastal dike [8]. Other researchers have studied probability-interval hybrid uncertainty problems in which for some parameters only their variation intervals can be obtained (due to the lack of, or poor quality of, samples). A comprehensive review of such a concept is provided in [9] including uncertainty modeling, uncertainty propagation analysis, structural reliability analysis, and reliability-based design optimization.

Developing a proper probabilistic model that is capable of estimating the quantity of interest (QoI) requires incorporating both types of uncertainty. This is also true in the probabilistic risk assessment of engineering systems [10, 11], where the probability of failure, $P_f$, needs to be determined.

The efficiency of probabilistic performance assessment depends on the sampling technique of the input random variables (RVs). Many of the finite element analysis packages act like a "black box" and calculate the QoI by spending extensive computational time [12]. If there were an analytical relationship between input and output variables, there would be no need for uncertainty quantification, as a simple response surface meta-model can also estimate the results.

Sampling is the main challenge in quantification of the impact of the material uncertainty. Although the Crude Monte Carlo Simulation (CMCS) is conventionally used to model probability distributions of QoIs, it has a major deficiency that makes it impractical for complex systems dealing with a transient analysis: CMCS requires vast amounts of simulations to provide a stable/reliable estimate of QoI (or $P_f$). Possible solutions to address this drawback are based on using efficient techniques such as:

- Pseudo-Random method: Latin Hypercube Sampling (LHS) guarantees samples to be drawn over the whole range of the distribution [15]. Given a system with RVs, $\mathbf{Z} = (z_1, z_1, ..., z_n)$ and corresponding distributions $D_1, D_1, ..., D_n$, first the range of each variable is split into $m$ non-overlapping intervals of equal marginal probability $1/m$. Then, sampling starts with the random selection of an interval followed by another random selection of a point inside it. The procedure is repeated until all intervals have been accessed, and none of them more than once. It is repeated for each of the $n$ RVs. One may note that the term pseudo is due to its dependence on pseudo-random numbers. These sequences are not truly random, because they are completely determined by an initial value, called the pseudo-random number generator's seed.

- Quasi-Random method: These random samples are those generated from a completely deterministic low-discrepancy process and have no inherent statistical properties. Two famous sequences are Halton [16] (which is defined by bases of prime numbers. Each dimension requires a unique prime number as a base) and Sobol [17] (which utilizes the base of two to construct finer uniform partitions of the unit interval and then reorders the coordinates in each dimension. It looks almost like a grid in lower dimensions but forms a lower discrepancy pattern at higher dimensions).

## 2. Hybrid Uncertainty Quantification: Objectives and Challenges

In many Civil Engineering simulations, the epistemic uncertainty roots in material properties, while the aleatory uncertainty originates from stochastic mechanical loads; see Fig. 1. Material characteristics are often presented based on probability density functions - PDFs - (e.g. normal and log-normal), while the stochastic loads might be simulated based on PDFs or extracted in a discrete form from hazard analysis (e.g., seismic hazard [13] and flood hazard [14]).

Overall, an engineering problem dealing with epistemic (e.g., material variability) and aleatory (e.g., loading variability) uncertainties can be schematically solved as shown in Fig.1. First appropriate realizations for both uncertainty types are sampled, and a detailed numerical model is developed. Next, the two sources of uncertainty should be combined and implemented using the numerical model. One simple technique to combine the two sources of uncertainties is to use Cloud analysis [18] (See the third row in Fig.1). In this method, the same number of realizations are required in epistemic and aleatory uncertainties to combine them randomly as shown in third row – left plot. Although this technique requires a small number of simulations, it may lead to a considerable bias. Since the number of samples is limited, Cloud analysis is prone to several bad combinations (i.e. severe loading with weak material/structure or vice versa) occur simultaneously in the system that can overly influence the mean and standard deviation.

Another technique is to combine each realization of aleatory uncertainty with all realizations from epistemic one. In this technique which is called Extended Cloud analysis [19], all the possible combinations are accounted for. However, the major drawback of this technique is the need for a large number of simulations [5].

Finally, the results of computer simulations can be used to develop the fragility function, and the response surface meta-models.

## 3. Estimating Quantities of Interest with Partial Information

A new improved method is discussed for predicting the entire set of output variables or quantities of interest from a small subset of simulations. This is important because we can reduce the computational cost. The main algorithmic contributions of our proposed method are twofold. First, we develop a sampling technique for choosing a subset of representative simulations, which allows us to improve the estimation accuracy. The main intuition is that not all simulations are equally important as a specific combination of ground motions and material conditions may lead to a more informative simulation compared to other combinations. This is because similar ground motions (with similar time, frequency, and energy content) and similar material combinations (because of random selection) might exist in the sampling seed. Therefore, our method aims to extract critical information on the intrinsic structure of uncertainties for the problem at hand. The second contribution is to develop a hybrid method that merges the results of the first step, known as matrix completion, with a regression model trained on the available partial simulations. Hence, our method not only looks for correlations between the output variables (e.g., displacements) but also takes into consideration important correlations between the input variables (i.e., ground motion and material uncertainties in this work) and the quantities of interest.

In the following, we discuss three main topics in detail: (1) matrix completion to recover unknown entries of an output matrix, (2) improved sampling using unsupervised machine learning, and (3) a new hybrid method for merging results from matrix completion and regression analysis.

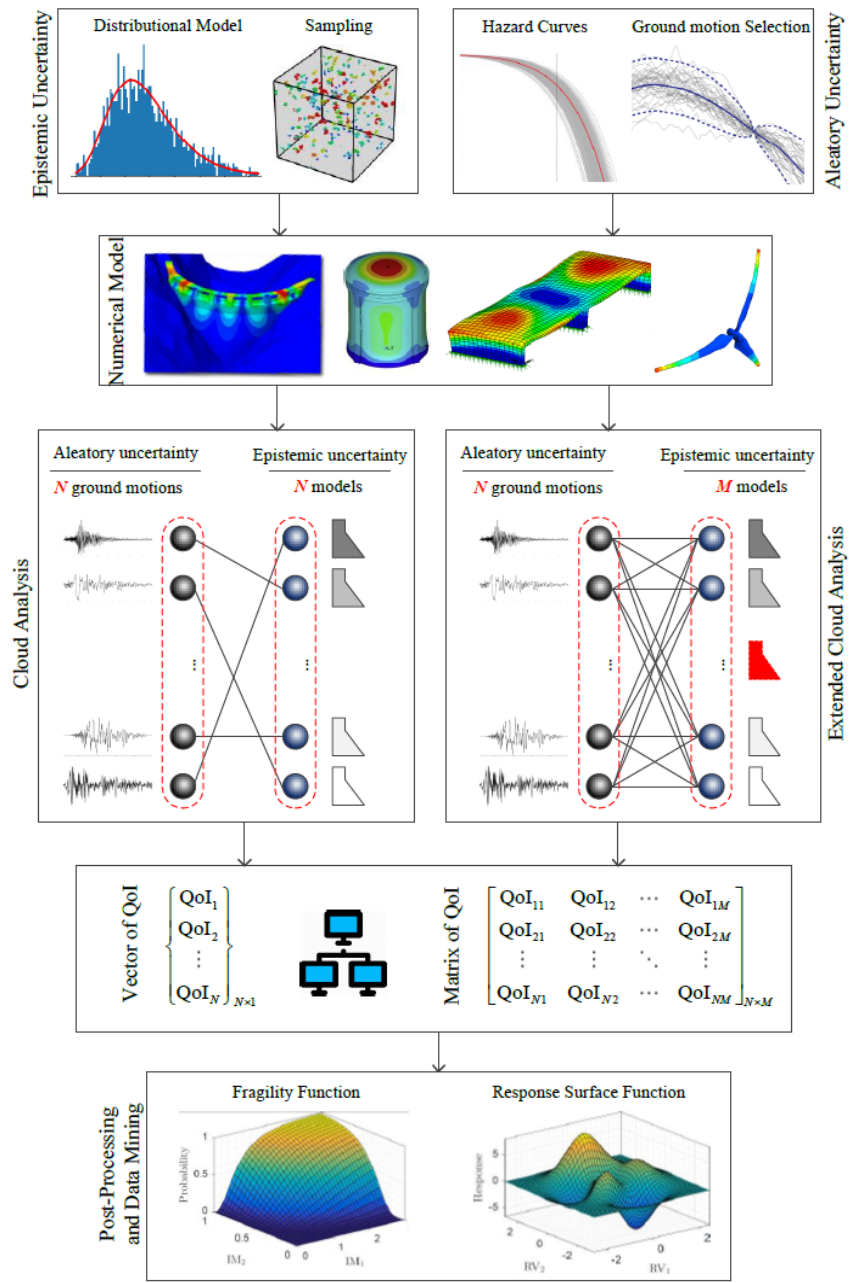

Fig. 1 – The general algorithm for Cloud analysis-based uncertainty quantification including aleatory and epistemic uncertainties [24]

## 3.1 Matrix Completion

Consider a matrix of outputs $X$ that has $N$ rows and $M$ columns. In this work, $N$ refers to the number of ground motion signals and $M$ represents the number of material combinations. To find all entries of the matrix $X$, one should perform $N*M$ simulations. However, we are interested in estimating the entire matrix $X$ using a subset of analyses to reduce computational costs and improve the overall efficiency. To achieve this goal, we focus on a setting where we perform only a fraction of simulations for each material combination. In other words, given a compression ratio parameter ($0<CR<1$) and a subset of entries with size $N*CR$ in each column of $X$, we want to predict the values of other missing entries. This problem is often referred to as matrix completion in the signal processing and machine learning literature [25].

The previous research [26] has proposed a mathematical optimization technique to use correlations between output variables for solving the matrix completion problem. In many practical problems, the matrix we wish to recover is approximately low-rank, which implies that the number of degrees of freedom is less than the ambient dimension. To exploit such a structure, we can approximate the matrix $X$ by a product of two smaller matrices $A$ and $B$, i.e., $X \sim A*B$, where $A$ has $N$ rows and $R$ columns, and $B$ has $R$ rows and $M$ columns. As a result, this decomposition allows us to recover the entire $N*M$ entries of the matrix $X$ from the following entries:

$$N \times R + R \times M = R \times (N + M), \quad where \quad R < \min(N, M) \tag{1}$$

Therefore, matrix completion aims to recover all $N*M$ entries of the matrix by first estimating the two matrices $A$ and $B$ with the total of $R*(N + M)$ entries. Note that for small values of $R$, the required number of entries $R*(N + M)$ is much less than $N*M$. The optimization problem can be cast as minimizing the distance between $X$ and $A*B$, often measured as the averaged Euclidean distance between corresponding entries of both matrices, i.e., comparing the entry in the $i$th row and $j$th column of $X$ with the entry of $A*B$ at the same location.

An important question that is often left unanswered is how to find the most informative entries from each column of X, so we achieve high estimation accuracies. A simple approach is to ignore the structure of input variables and choose $N*CR$ entries from each column of $X$ uniformly at random. The main disadvantage of uniform sampling is that we may miss critical information about the structure of the matrix $X$. For example, it is very likely to miss a few entries of $X$ that correspond to rare events, i.e., major structural damage in our application. Thus, a challenging problem that we address in this work is related to finding informative sampling techniques to improve the quality of matrix completion.

## 3.2 Improved Sampling via Cluster Analysis for Matrix Completion

This section proposes a new sampling technique for selecting a subset of informative simulations to cover distinct combinations of input variables. The introduced method employs a clustering technique to find groups of similar input variables. In machine learning and signal processing, clustering is a fundamental problem that has received great attention. For example, clustering techniques can be used as exploratory tools to uncover important structures in a high-dimensional parameter space.

To be formal, we propose to use the $k$-medoids clustering algorithm in the parameter space of ground motions. This algorithm partitions the $N$ ground motion signals into $k$ groups, and the number of $k$ clusters assumed to be known a priori:

$$obtain \quad k \; clusters \quad \longleftarrow k - medoids(g_1, g_2, ..., g_N) \tag{2}$$

where $g_1, ... g_N$ are $N$ ground motions. Thus, the clustering step allows us to automatically identify various groups of similar ground motions and we can employ uniform sampling within each cluster. Consequently, the overall sampling technique is not uniform anymore because we often encounter groups of varying sizes. For example, the group of ground motions corresponding to rare events is typically much smaller than the other clusters.

Note that after clustering the input variables, we should use our sampling technique $M$ times as we want to find $N*CR$ entries from each column of $X$. Hence, this sampling strategy enables us to collect a diverse set of simulations for predicting the remaining entries, as explained in the previous section. A further advantage of using $k$-medoids clustering is that we can use different types of distance functions to represent the pairwise similarities between ground motions.

## 3.3 Incorporating Regression Analysis into Matrix Completion

Another contribution of this work is to further improve the performance of matrix completion using other machine learning techniques. Particularly, we propose to train a regression model to find the input-output relationships in addition to the matrix completion step, which only considers correlations between output variables. Hence, we will have two predicted values for each unknown entry of $X$ to obtain more accurate predictions. The proposed method falls into ensemble learning, which is an active area of research in machine learning with promising results to improve the performance of many existing techniques.

To discuss our method in more detail, recall that we have access to a total of $N*CR*M$ entries of $X$, where $CR<1$. Using this information, we train a regression model such as support vector regression to extract the input-output relationship. In this case, the input parameter space includes information about both ground motions and material uncertainties. After learning the mapping between inputs and desired outputs, we predict the quantities of interest for remaining entries one by one. This will give us one estimate of the entire matrix $X$, which we refer to as $X_{est(2)}$. Next, we compute the average of our previous estimates with the new predictions using the regression model. Let $X_{est(1)}$ be the estimated matrix of outputs obtained via matrix completion. Then, the final estimate can be computed in the following form:

$$X_{est} = \frac{1}{2}\left(X_{est(1)} + X_{est(2)}\right) \tag{3}$$

In this work, we considered training one regression model in addition to the matrix completion step. However, an interesting future research direction is to use T regression models [27] (e.g., linear and non-linear techniques) and take the average of all estimated matrices to achieve more accurate predictive models.

## 4. Case Study Description

A hypothetical telecommunication tower (with the height > 400 m) is assumed as case study. The tower is assumed to be a reinforced concrete (RC) structure. The concrete service core is the main load-carrying structure of the tower that transfers the lateral and gravitational loads to the foundation.

As mentioned in the previous section, a large number of transient analyses are performed in this paper. On the other hand, modeling aspects such as material nonlinearities, i.e. cracking, crushing, and damage, and geometric nonlinearities, i.e. P-Delta effects, and large displacements, as well as interaction between the different structural components, should be considered. As a consequence, performing 3D finite element analysis to address the transient effects together with the other nonlinearity issues becomes very time consuming. It is desirable to look for a model that not only requires fewer elements and less computational time, but also provides the desired outputs with an acceptable loss of accuracy. One of the best alternatives is to model the structure using uniaxial fiber beam-column elements.

A 2D nonlinear model of the tower, including the head structure, shaft and transition, is developed using OpenSees [20]. The service core is modeled using 2D force-based nonlinear fiber beam-column elements [28] with five integration points along their length. The core cross-section is discretized into concrete and steel fibers. The superstructure is idealized using equivalent mass of the floors. The base elevation of the building is constrained in the lateral and rotational degrees of freedom, excluding the effects of soil-structure interaction [21].

The concrete is modeled based on the uniaxial Kent-Scott-Park constitutive model [29] with degraded linear unloading/reloading stiffness as shown in Fig. 2a. In this model $f'_c$ is the compressive strength of concrete and $\varepsilon_0$ is the strain at the peak strength. In addition, $f_u$ and $\varepsilon_u$ are the ultimate compressive stress and its corresponding stain. Steel is modeled based on the Giuffre-Menegotto-Pinto model with isotropic strain hardening (Fig. 2b), in which a transition curve is defined to avoid the unsmoothed response of bilinear kinematic hardening behavior at the yield point, and consequently, the path-dependent nature of the material can be traced effectively. Furthermore, the Bauschinger effect is intrinsically defined in the material stress-strain curve so that the deterioration of strength in the element behavior is automatically modeled. The yield strength of the steel is 400 MPa and 1% post-yield stiffness is assumed to account for the hardening of steel beyond the yield strength.

In this study, Rayleigh damping is adopted where the mass proportional part is constant during the analysis; however, the stiffness proportional part changes according to the updated stiffness matrix of the structure. In other words, the damping property is allowed to vary and be updated for each load step of the transient analysis.

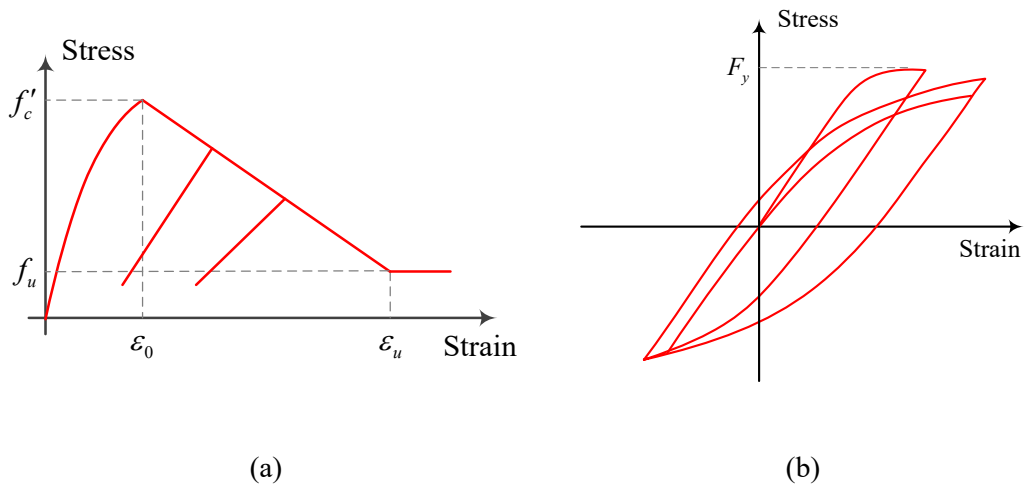

(a)                                            (b)

Fig. 2 –Stress-strain relationship for (a) concrete, (b) steel [21]

Since both material and ground motion record-to-record uncertainties are studied in this paper, multiple assumptions are made to these models. In the numerical models, the Gaussian distributional model is used. A total of 10 random models are generated using Monte Carlo Simulation with LHS, to consider the variability in 18 material parameters:

- Concrete: weight per volume, compressive strength, strain at maximum strength, crushing strength, strain at crushing strength

- Steel: yield strength, initial elastic tangent, strain-hardening ratio, area of bars

- System: Damping ratio

A total of 100 ground motions are selected. They are randomly chosen from the PEER database to cover a wide range of potential ground motions. This is in line with Cloud analysis, where a large dataset of unscaled (as-recorded) signals are applied to the structural system [30]. For each ground motion, 31 intensity measure (IM) parameters are extracted to develop a side information matrix [23]. These 31 IM parameters are selected from a comprehensive list of over 70 IM parameters found in Hariri-Ardebili and Saouma [22]. A few important IM parameters are listed below as examples: 1) acceleration / velocity / displacement spectrum intensity, 2) effective peak acceleration / velocity, 3) spectral acceleration / velocity / displacement in first five modes, 4) PGA / PGV / PGD, 5) root mean square of acceleration / velocity / displacement, 6) Arias intensity and significant duration, and 7) cumulative absolute velocity / displacement

## 5. Results

Next, 100 ground motions are combined with 10 combined samples from material uncertainty, and a total of 1,000 nonlinear transient simulations are performed. Results are stored in two main matrices of size 100*10: one for maximum top displacement and another for maximum base shear (red and yellow matrix in Fig. 3). These are the main EDP matrices, while we have two side matrices of size 100*31 (light blue matrix in Fig.

3) obtained from ground motion post-processing, and 10*18 (dark blue matrix in Fig. 3) obtained from initial material uncertainty.

In Fig. 3, we assume that only a portion of the entire simulations are already available (yellow cells), while the others are unknown/missing (red cells). The idea is to predict the missing cells and reconstruct the entire matrix (the purple matrix in Fig. 3). Depending on the number of the initial available data, the errors associated with ground motions, materials, or the combined error can be defined.

A parametric study is then performed, assuming only a portion of the original matrix. A Compression Ratio (CR) from [0.1:0.1:0.5] is selected. One may note that CR = 0.2 (for example) means that only 20% of the initial data (in our case only 200 analyses out of 1000) is known, and the remaining cells are assumed to be empty. Since the process of selection of initial "known" data is random, 50 trials are performed for each one to reduce the randomness in the data section process. As discussed before, three sampling techniques are compared:

- Uniform sampling,
- New (proposed) sampling,
- New (proposed) sampling plus regression (ensemble learning).

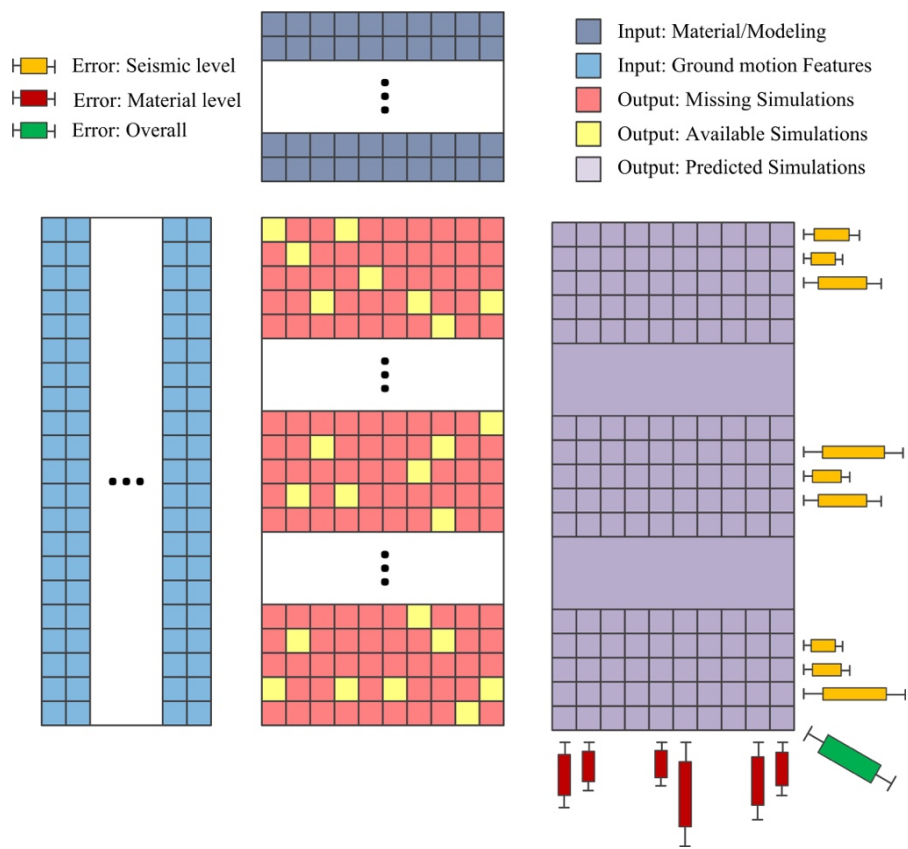

Fig.3 – Schematic presentation of input-output data and matrix completion process; note: Error is shown using error bar plots

Fig. 4 presents the final results of matrix completion for both displacement and base shear responses. In this figure, the horizontal axis is different compression ratios, while the vertical one shows the normalized estimation error for different techniques. The following major conclusions can be drawn:

- The general trend of predicted error for both engineering demand parameters is similar. This means that there is a relatively high correlation between displacement and base shear.

- For both EDPs, increasing CR reduces the mean normalized estimation error. This is qualitatively intuitive as increasing CR provides more inputs to the prediction algorithm. However, one may decide (quantitatively) an acceptable threshold for CR.

- The normalized estimation error reduction pattern in both proposed algorithms is linear in semi-logarithmic scale, while is nonlinear for uniform sampling.

- The proposed sampling algorithm with regression outperforms the other methods.

- The proposed sampling algorithm plus regression can predict the entire matrix with 50% accuracy having only 10% of the initial data. It increases to 90% accuracy having 50% of the initial data. This method then saves lots of computational time which is valuable especially for huge finite element models.

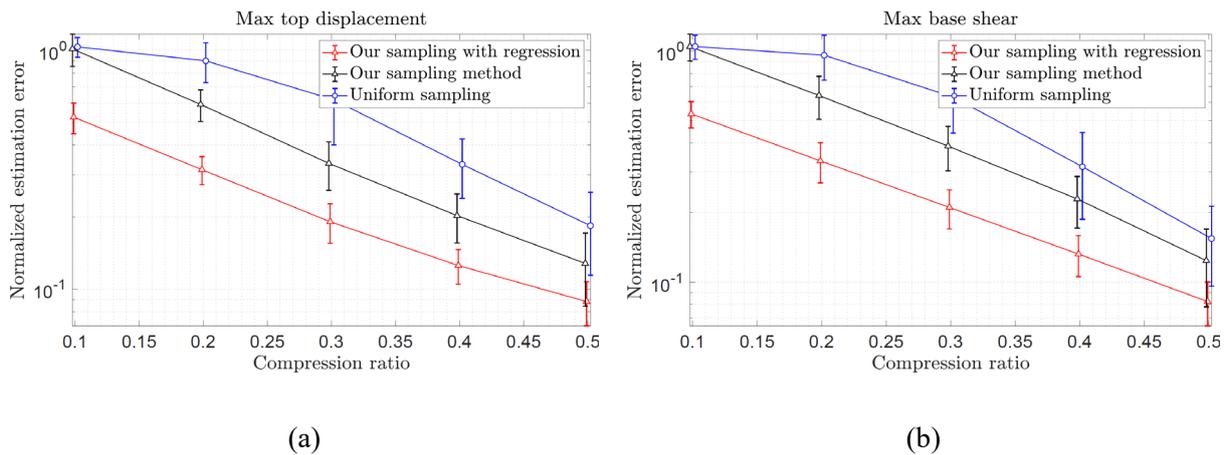

(a)                                                                    (b)

Fig.4 – Results of matrix completion with varying compression ratios for: (a) displacement, (b) base shear.

## 6. Conclusions

This paper presented an improved algorithm for reducing the number of the nonlinear seismic analyses required to quantify the impact of hybrid epistemic and aleatory uncertainties in an effective manner. Three matrix completion methods are proposed and applied on a case study structure. The method is based on predicting the entire set of output variables or quantities of interest from a small subset of initial simulations. The main algorithmic contributions of our proposed method are twofold. First, we develop a sampling technique for choosing a subset of representative simulations, which allows us to improve the accuracy of response estimation. The main intuition is that not all simulations are equally important as a specific combination of ground motions and material conditions may lead to a more informative simulation compared to other combinations. The second contribution is to develop a hybrid method that merges the results of the first step with a regression model trained on the available partial simulations.

An improved sampling method using unsupervised machine learning is combined with matrix completion and regression analysis to predict the structural responses (i.e., displacements and base shear) of a multi-degree-of-freedom system. Results show that the proposed algorithm can effectively (in terms of accuracy and cost of analyses) estimate the full set of nonlinear simulations by having only a small portion of all the required simulations (with a reasonable amount of error – Fig 4 - depending on the application).

# 7. References

[1] Der Kiureghian A, Ditlevsen O (2009): Aleatory or epistemic? Does it matter? *Structural safety*, *31*(2), 105-112.

[2] Koduru SD (2015, July): Strategies for separation of aleatory and epistemic uncertainties. In *12th International Conference on Applications of Statistics and Probability in Civil Engineering (ICASP12)*, Vancouver, Canada.

[3] Ross JL, Ozbek MM, Pinder GF (2009): Aleatoric and epistemic uncertainty in groundwater flow and transport simulation. *Water Resources Research*, *45*(12).

[4] Celik OC, Ellingwood BR (2010): Seismic fragilities for non-ductile reinforced concrete frames–Role of aleatoric and epistemic uncertainties. *Structural Safety*, *32*(1), 1-12.

[5] Dolšek M (2012): Simplified method for seismic risk assessment of buildings with consideration of aleatory and epistemic uncertainty. *Structure and infrastructure engineering*, *8*(10), 939-953.

[6] Hariri-Ardebili MA, Saouma VE (2016): Collapse fragility curves for concrete dams: comprehensive study. *Journal of Structural Engineering*, *142*(10), 04016075.

[7] Merz B, Thieken AH (2005): Separating natural and epistemic uncertainty in flood frequency analysis. *Journal of Hydrology*, *309*(1-4), 114-132.

[8] Helton JC, Johnson JD, Oberkampf WL, Sallaberry CJ (2010): Representation of analysis results involving aleatory and epistemic uncertainty. *International Journal of General Systems*, *39*(6), 605-646.

[9] Jiang C, Zheng J, Han X (2018): Probability-interval hybrid uncertainty analysis for structures with both aleatory and epistemic uncertainties: a review. *Structural and Multidisciplinary Optimization*, *57*(6), 2485-2502.

[10] Winkler RL (1996): Uncertainty in probabilistic risk assessment. *Reliability Engineering & System Safety*, *54*(2-3), 127-132.

[11] Stewart M, Melchers RE (1997): *Probabilistic risk assessment of engineering systems*. Springer.

[12] Ramu M, Prabhu RV (2013): METAMODEL BASED ANALYSIS AND ITS APPLICATIONS: A REVIEW. *Acta Technica Corviniensis-Bulletin of Engineering*, *6*(2).

[13] McGuire RK (1995): Probabilistic seismic hazard analysis and design earthquakes: closing the loop. *Bulletin of the Seismological Society of America*, *85*(5), 1275-1284.

[14] National Research Council (1988): Estimating probabilities of extreme floods: methods and recommended research. National Academies.

[15] Iman RL, Conover WJ (1982): A distribution-free approach to inducing rank correlation among input variables. *Communications in Statistics-Simulation and Computation*, *11*(3), 311-334.

[16] Halton JH (1964): Algorithm 247: Radical-inverse quasi-random point sequence. *Communications of the ACM*, *7*(12), 701-702.

[17] Sobol' IYM (1967): On the distribution of points in a cube and the approximate evaluation of integrals. *Zhurnal Vychislitel'noi Matematiki i Matematicheskoi Fiziki*, *7*(4), 784-802.

[18] Jalayer F (2003): Direct probabilistic seismic analysis: implementing non-linear dynamic assessments. Stanford, CA: Stanford University.

[19] Hariri-Ardebili MA, Saouma VE (2017): Single and multi-hazard capacity functions for concrete dams. *Soil Dynamics and Earthquake Engineering*, *101*, 234-249.

[20] McKenna F, Fenves GL, Scott MH, Jeremic B (2000): Open System for Earthquake Engineering Simulation (OpenSees), Pacific Earthquake Engineering Research Center, University of California, Berkeley, CA.

[21] Hariri-Ardebili MA, Rahmani-Samani H, Mirtaheri M (2014): Seismic stability assessment of a high-rise concrete tower utilizing endurance time analysis. *International Journal of Structural Stability and Dynamics*, *14*(06), 1450016.

[22] Hariri-Ardebili MA, Saouma VE (2016): Probabilistic seismic demand model and optimal intensity measure for concrete dams. *Structural Safety*, *59*, 67-85.

[23] Hariri-Ardebili MA, Barak S (2020): A series of forecasting models for seismic evaluation of dams based on ground motion meta-features. *Engineering Structures*, *203*, 109657.

[24] Hariri-Ardebili MA, and Pourkamali-Anaraki F (2019): Matrix completion for cost reduction in finite element simulations under hybrid uncertainties. *Applied Mathematical Modelling*, 69, 164-180.

[25] Candès EJ, and Benjamin R (2009): Exact matrix completion via convex optimization. *Foundations of Computational mathematics* 9(6), 717.

[26] Jain P, Praneeth N, Sujay S. (2013): Low-rank matrix completion using alternating minimization. In Proceedings of the forty-fifth annual ACM symposium on Theory of computing, pp. 665-674.

[27] Hastie T, Tibshirani R, Friedman J (2009): The elements of statistical learning: data mining, inference, and prediction. Springer Science & Business Media.

[28] Scott MH, and Fenves GL. (2006): Plastic hinge integration methods for force-based beam–column elements. *Journal of Structural Engineering* 132(2), 244-252.

[29] Scott BD, Park, R, Priestley MJN (1982): Stress-Strain Behavior of Concrete Confined by Overlapping Hoops at Low and High Strain Rates. *ACI Journal Proceedings*, 79, 13-27.

[30] Jalayer F, De Risi R, Manfredi G (2015): Bayesian Cloud Analysis: efficient structural fragility assessment using linear regression. *Bulletin of Earthquake Engineering* 13, 1183-1203.